\documentclass[aps,pra,floatfix,superscriptaddress,twocolumn,amssymb]{revtex4}
\usepackage{amsmath}
\usepackage{mathrsfs}
\usepackage{graphicx}
\usepackage{dcolumn}
\usepackage{bm}
\usepackage{euscript}
\usepackage{simon}

\begin{document}
\title{A laser frequency comb that enables radial velocity measurements with a precision of 1 cm s$^{-1}$}
\author{Chih-Hao Li}
\affiliation{Harvard-Smithsonian Center for Astrophysics, Cambridge
Massachusetts} \affiliation{Department of Physics, Harvard
University, Cambridge, Massachusetts}

\author{Andrew J. Benedick}
\affiliation{Department of Electrical Engineering and Computer
Science and Research Laboratory for Electronics, Massachusetts
Institute of Technology, Cambridge, Massachusetts}

\author{Peter Fendel}
\affiliation{Department of Electrical Engineering and Computer
Science and Research Laboratory for Electronics, Massachusetts
Institute of Technology, Cambridge, Massachusetts}
\affiliation{MenloSystems Inc., Newton, New Jersey}

\author{Alexander G. Glenday}
\affiliation{Department of Physics, Harvard University, Cambridge,
Massachusetts}

\author{Franz X. K${\rm \ddot{a}}$rtner}
\affiliation{Department of Electrical Engineering and Computer
Science and Research Laboratory for Electronics, Massachusetts
Institute of Technology, Cambridge, Massachusetts}

\author{David F. Phillips}
\affiliation{Harvard-Smithsonian Center for Astrophysics, Cambridge
Massachusetts}

\author{Dimitar Sasselov}
\affiliation{Harvard-Smithsonian Center for Astrophysics, Cambridge
Massachusetts}

\author{Andrew Szentgyorgyi}
\affiliation{Harvard-Smithsonian Center for Astrophysics, Cambridge
Massachusetts}

\author{Ronald L. Walsworth}
\affiliation{Harvard-Smithsonian Center for Astrophysics, Cambridge
Massachusetts} \affiliation{Department of Physics, Harvard
University, Cambridge, Massachusetts}

\date{\today}
\maketitle

{\bf Searches for extrasolar planets using the periodic Doppler
shift of stellar spectral lines have recently achieved a precision
of $60~cm~s^{-1}$ (ref 1), which is sufficient to find a
5-Earth-mass planet in a Mercury-like orbit around a Sun-like star.
To find a 1-Earth-mass planet in an Earth-like orbit, a precision of
$\sim 5~cm~s^{-1}$ is necessary. The combination of a laser
frequency comb with a Fabry-P${\bf \rm \acute{e}}$rot filtering
cavity has been suggested as a promising approach to achieve such
Doppler shift resolution via improved spectrograph wavelength
calibration$\bf ^{2-4}$, with recent encouraging results$\bf ^5$.
Here we report the fabrication of such a filtered laser comb with up
to 40-$GHz$ ($\bf \sim 1$-${\bf \AA}$) line spacing, generated from
a 1-$GHz$ repetition-rate source, without compromising long-term
stability, reproducibility or spectral resolution. This
wide-line-spacing comb, or `astro-comb', is well matched to the
resolving power of high-resolution astrophysical spectrographs. The
astro-comb should allow a precision as high as $1~cm~s^{-1}$ in
astronomical radial velocity measurements.}

The accuracy and long-term stability of state-of-the-art
astrophysical spectrographs are currently limited by the
wavelength-calibration source$^{6,7}$, typically either
thorium-argon lamps or iodine absorption cells$^8$. In addition,
existing calibration sources are limited in the red-to-near-IR
spectral bands most useful for exoplanet searches around M stars$^9$
and dark matter studies in globular clusters$^{10}$. Iodine cells
have very few spectral lines in the red and near-IR spectral bands,
while thorium-argon lamps have limited lines and unstable bright
features that saturate spectrograph detectors. Recently, laser
frequency combs$^{11}$ have been suggested as potentially superior
wavelength calibrators$^{2,3}$ because of their good long-term
stability and reproducibility, and because they have useful lines in
the red-to-near-IR range. The absolute optical frequencies of the
comb lines are determined by $f = f_{ceo} + m \times f_{rep}$, where
$f_{rep}$ is the repetition rate, $f_{ceo}$ is the carrier-envelope
offset frequency and $m$ is an integer. Both $f_{rep}$ and $f_{ceo}$
can be synchronized with radio-frequency oscillators referenced to
atomic clocks. For example, using the generally available Global
Positioning System (GPS), the frequencies of comb lines have
long-term fractional stability and accuracy of better than
$10^{-12}$. For the calibration of an astrophysical spectrograph,
fractional stability and accuracy of $3 \times 10^{-11}$ are
sufficient to measure a velocity variation of $1~cm~s^{-1}$ in
astronomical objects. In addition, using GPS as the absolute
reference allows the comparison of measurements at different
observatories.

For existing laser combs, $f_{rep}$ is usually $<1~GHz$ (ref.~12),
which would require a spectrograph with a resolving power of $R =
\lambda / \delta \lambda  \gg 10^5$  to resolve individual comb
lines (here $\delta \lambda$ is the smallest difference in
wavelengths that can be resolved at wavelength $\lambda$). In
practice, astrophysical spectrographs tend to have a resolving power
of $R \sim 10^4-10^5$ owing to physical limitations on the
instruments, including the telescope aperture, the grating
collimator diameter and the grating blaze. Thus, a laser comb must
have line spacing $>10~GHz$ to serve as a practical wavelength
calibrator. Therefore, we augmented a 1-$GHz$-repetition-rate laser
comb with a stable broadband Fabry-P${\rm \acute{e}}$rot (FP) cavity
to increase the comb line spacing to $40~GHz$ over a range
$>1,000~{\rm \AA}$. This novel$^2$, wide-line-spacing `astro-comb'
can provide improved wavelength calibration for a wide range of
existing and planned astrophysical spectrographs.

The astro-comb set-up is shown schematically in Fig.~1. An
octave-spanning optical frequency comb with a 1-$GHz$ repetition
rate (`source-comb') is generated by a mode-locked Ti:sapphire
femtosecond laser$^{13}$. The linewidth of each comb line is
$<1~kHz$, with both $f_{rep}$ and $f_{ceo}$ stabilized using
low-noise frequency synthesizers, which can be referenced to an
atomic clock. The stabilized source-comb light passes through an FP
cavity that filters out unwanted comb lines and increases the line
spacing. The FP cavity is stabilized by an injected diode laser
signal that is itself stabilized to the Rb D1 line (7,947~${\rm
\AA}$) using a dichroic-atomic-vapour laser lock (DAVLL$^{14}$).

\begin{figure*}
\includegraphics[width=7.2in]{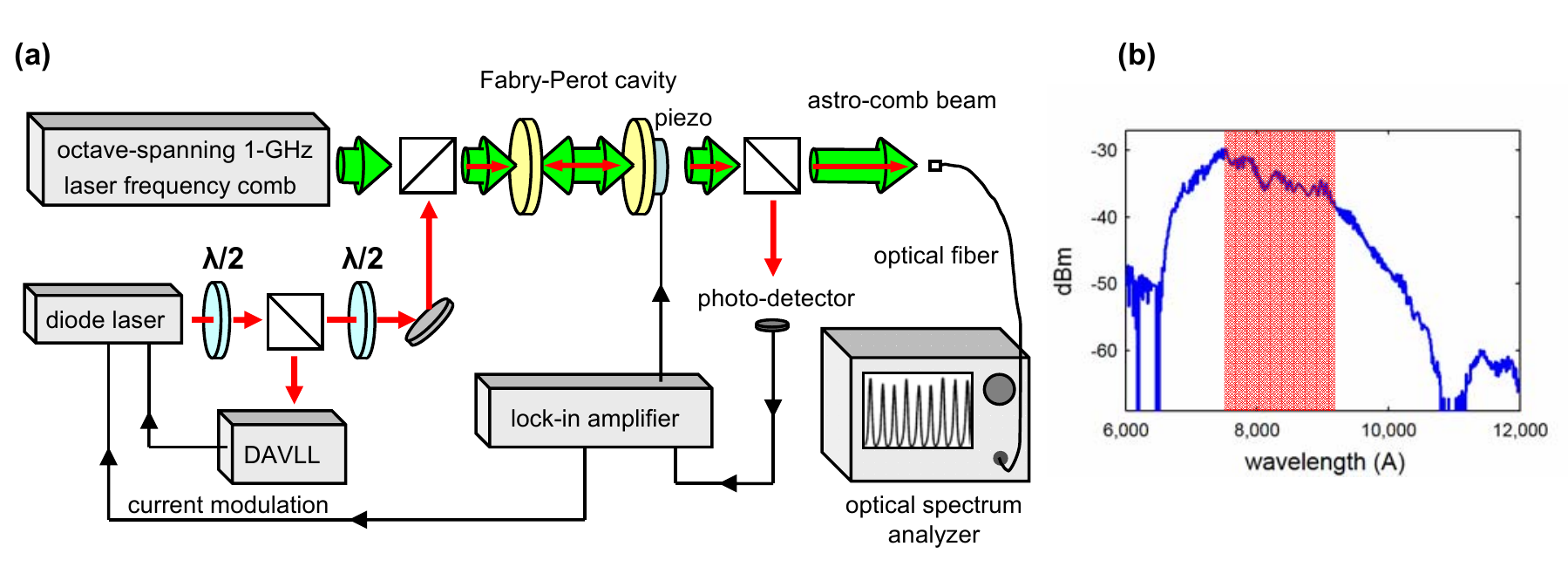}
\caption{{\bf Block diagram of the astro-comb. a}, A stabilized
1-$GHz$ frequency comb (`source-comb') using a mode-locked
femtosecond laser passes through an FP cavity that filters out
unwanted comb lines and increases the line spacing to at most
40~$GHz$ ($\sim 1~{\rm \AA}$). For the demonstration spectra shown
herein this paper, the output beam from the astro-comb is collected
by a single-mode fibre and measured using an optical spectrum
analyzer (Ando 6317) with resolution $\sim 8~GHz$ and
reproducibility $2~GHz$. {\bf b}, Output spectrum of the 1-$GHz$
source-comb. Typical operating parameters of the source-comb$^{13}$
are $600~mW$ of output power and an output spectrum from $6,000~{\rm
\AA}$ to $12,000~{\rm \AA}$ with $9.3~W$ of pump power. The shaded
area is the spectral range in which the current FP cavity mirrors
have small GDD and hence provide good suppression of extraneous comb
lines. The quantity dBm is ten times the logarithm of the power
referenced to $1~mW$.}
\end{figure*}

To realize an astrophysical wavelength calibrator, the FP cavity
must filter comb lines over a broad spectral range. The mirrors used
in the plane-parallel FP cavity have ~99$\%$ reflectivity and
optimized group delay dispersion (GDD) ($<10~fs^2$) in the range of
7,700~${\rm \AA}$ to 9,200~${\rm \AA}$. We measured the finesse of
the FP cavity to be ~250 at 7,947~${\rm \AA}$, which is consistent
with the theoretical limit estimated from the mirror reflectivity
and Fresnel losses. The GDD-optimized mirrors enable the generation
of a filtered comb spanning a bandwidth of $\sim 1,000~{\rm \AA}$.
With straightforward adjustment of the free spectral range of the FP
cavity to approximately equal an integer multiple of $f_{rep}$, we
realized such comb-line filtering. For example, Fig.~2 shows the
measured astro-comb output spectrum spanning a bandwidth of $\sim
1,000~{\rm \AA}$, with 37-$GHz$ line spacing and power $\sim
10-100~nW$ in each comb line. If the ratio of free spectral range to
$f_{rep}$ is not an integer, the span of the filtered comb lines is
narrower and groups of filtered comb lines appear repeatedly within
the bandwidth of the mirrors (Fig.~2c). This `Vernier-like' pattern
can be shifted in wavelength and modified by varying the source-comb
$f_{ceo}$ and $f_{rep}$ or the free spectral range of the FP cavity.
The adjustability of the astro-comb-line pattern may assist the
calibration of spectrographs over the bandwidth of the mirrors.

\begin{figure}
\includegraphics[width=3.5in]{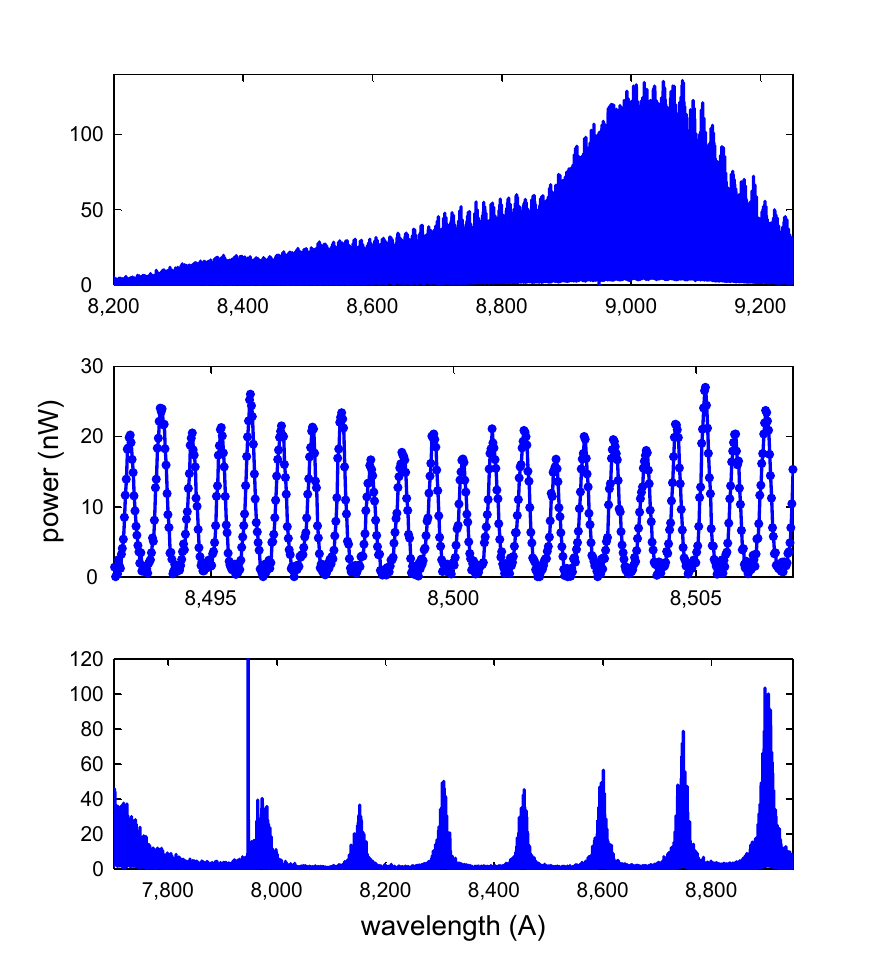}
\caption{{\bf Example astro-comb output spectrum with 37-$GHz$ line
spacing. a}, The astro-comb is tuned to span a bandwidth of
$1,000~{\rm \AA}$ . The resolution of the figure is not high enough
to show individual lines. The observed amplitude variation is
primarily due to the amplitude variation of the source-comb, with a
minor contribution from `mismatch' of the source-comb lines and the
transmission resonances of the FP cavity caused by the residual GDD
of the mirrors ( $0-5~fs^2$). Small line shifts due to residual
mirror GDD can be determined experimentally to high precision
($<1~cm~s^{-1}$). For the TiO$_2$/SiO$_2$ multi-layer mirrors used
in the FP cavity, line stability $\sim 1~cm~s^{-1}$ is expected on
timescales of several years. {\bf b}, A small portion of the full
output spectrum, showing individual filtered comb lines. The width
of the lines is set by the optical spectrum analyzer's resolution ($
8~GHz$). {\bf c}, Intentional mismatch between source-comb spacing
$f_{rep}$ and FP cavity free spectral range causes groups of
filtered comb lines to appear repeatedly within the bandwidth of the
mirrors. The prominent line at $7,947~{\rm \AA}$ is the injected
diode laser signal used to stabilize the FP cavity.}
\end{figure}

In addition to tunable line spacing up to $40~GHz$, appropriate for
use with astrophysical spectrographs, the astro-comb exhibits the
stability and extraneous-line suppression necessary in an improved
wavelength calibrator. By comparisons with a hydrogen maser, we
determined the frequency fractional stability of the source-comb
(characterized by $f_{rep}$ and $f_{ceo}$) to be better than
$10^{-12}$ on timescales of seconds to hours. Ideally, the FP cavity
changes only the amplitude of the astro-comb's output lines, and not
their frequency. Thus, the required stability of the FP cavity is
much less stringent than that of the source-comb. However, several
source-comb lines lie inside the resolution bandwidth of a typical
astrophysical spectrograph. Although the FP cavity has finite
suppression of neighbouring comb lines, instability in it leads to
changes in the line shape of the astro-comb output spectrum as
measured by an astrophysical spectrograph. In Fig.~3, we show a
direct measurement of the suppression of extraneous lines of the
astro-comb. The measured single-sided suppression of extraneous comb
lines of more than 25~dB is consistent with the measured FP cavity
finesse of 250. The FP cavity is stabilized by locking one
transmission resonance maximum to a DAVLL-stabilized diode laser.
The DAVLL-stabilized FP cavity is quite robust, remaining locked for
periods of days. The absolute uncertainty in the DAVLL stabilized
system is below $0.5~MHz$, which is more than sufficient to maintain
a sensitivity of $1~cm~s^{-1}$. (As noted above, the required FP
cavity stability is much less stringent than the required 10-$kHz$
source-comb stability.) Residual frequency noise in the DAVLL is
$<300~kHz~Hz^{-1/2}$; corresponding to an amplitude fluctuation of
$<0.1\%$ for the suppressed (extraneous) comb lines. The resultant
frequency noise in the desired astro-comb line spacing is
$<3~kHz~Hz^{-1/2}$. Taking advantage of the ultrastable source-comb
lines, the astro-comb output spectrum measured by the spectrograph
is more stable than the FP cavity by more than two orders of
magnitude. Consequently, the stability of the astro-comb is more
than adequate for wavelength calibration of astrophysical
spectrographs to $1~cm~s^{-1}$ sensitivity.

\begin{figure}
\includegraphics[width=3.5in]{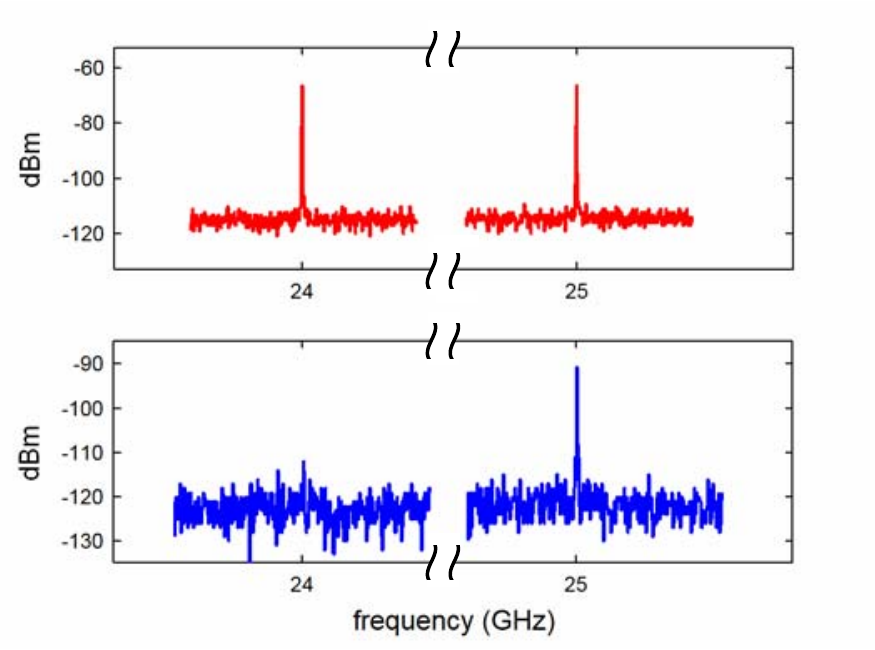}
\caption{{\bf Suppression of extraneous source-comb lines for the
astro-comb.} Here, the FP cavity is tuned such that the astro-comb
has a 25-$GHz$ line spacing. A fast photo-diode and spectrum
analyzer are used to measure the power in the 24-$GHz$ and 25-$GHz$
beatnotes from the 1-$GHz$ source-comb {\bf (a)} and the 25-$GHz$
astro-comb {\bf (b)} in the band $8,150-8,450~{\rm \AA}$. The 22-dB
suppression of the 24-$GHz$ beatnote in the astro-comb output signal
corresponds to a 25-dB single-sided suppression of extraneous comb
lines, consistent with the FP cavity finesse of 250. When the
astro-comb is used as a wavelength calibrator for an astrophysical
spectrograph, the extraneous-line suppression will be regularly
monitored.}
\end{figure}

In May 2008, we will deploy an astro-comb wavelength calibrator at
the Multiple Mirror Telescope (MTT) on Mt Hopkins, Arizona. We will
demonstrate the ability of the astro-comb to calibrate the
Hectochelle multi-object echelle spectrograph$^{15}$ in a 150-${\rm
\AA}$ bandwidth around $8,500~{\rm \AA}$, which will be especially
useful for the study of dark matter and other phenomena in globular
clusters. Here we estimate the expected wavelength calibration
precision of the astro-comb over a typical 10-hour MMT/Hectochelle
measurement. The 1-$GHz$ source-comb, referenced to GPS, will have
spectral lines with accuracy and long-term fractional stability
better than $10^{-12}$. The FP cavity will use mirrors (Lambda
Research) with 99$\%$ reflectivity and minimal GDD ($<1~fs^2$ over
the 150-${\rm \AA}$ bandwidth of one Hectochelle order). The free
spectral range of the FP cavity will be set to $\sim 25~GHz$, which
maximizes the calibration sensitivity$^2$. Residual FP cavity mirror
GDD and fluctuations of the FP cavity resonance$^{16}$ (width $\sim
150~MHz$) will lead to changes in the extraneous-line suppression,
which is typically $\sim 4 \times 10^{-3}$ (that is, 25~dB), of up
to 0.2$\%$. Therefore, an upper-limit estimate of the uncertainty of
astro-comb-line centres is $(4 \times 10^{-3}) \times 0.2\% \times
1~GHz \sim 8~kHz$. This uncertainty results in a systematic error in
astrophysical velocity measurements of approximately
$(8~kHz/377~THz) \times (3 \times 10^{10}~cm~s^{-1}) < 1~cm~s^{-1}$.
In practice, the precision of Doppler-shift/redshift measurements
will also be affected by telescope instability and astronomical
light-source fluctuations$^6$.

Beyond our first demonstration, astro-combs should enable many
observations that have previously been considered technically
unachievable. One example is the search for a 1-Earth-mass planet in
an Earth-like orbit around a Sun-like star, which requires a
sensitivity of $5~cm~s^{-1}$ and stability on at least a 1-year
timescale. In 2009 or 2010, we will deploy an astro-comb at the
HARPS-NEF (High-Accuracy Radial-velocity Planet Searcher of the New
Earths Facility) spectrograph ($R = 120,000$) being built by the
Harvard Origins of Life Initiative for the William Herschel
telescope to search for exoplanets. We will use broadband mirrors
with 99$\%$ reflectivity and optimal GDD in the FP cavity to
generate stable calibration lines in appropriate spectral bands,
which we expect will make the HARPS-NEF spectrograph sensitive
enough to find Earth-like planets. Additional wavelength coverage
can also be realized by frequency doubling the source-comb. Another
possible application of the astro-comb is the Sandage-Loeb
test$^{17,18}$, a direct measurement of the decelerating expansion
of the early Universe. This test requires an observation period of
$>10$ years with existing wavelength calibrators, but should be
feasible with an observation period of $\sim 3$ years using
astro-combs. Thus, by enabling a velocity-shift precision of $\sim
1~cm~s^{-1}$, broad wavelength coverage and reproducibility over
many years and between telescopes, astro-combs should revolutionize
astrophysical spectroscopy.

\section*{Acknowledgements}
This project is supported by the Harvard University Origins of Life
Initiative, the Smithsonian Institution, and DARPA.


\end{document}